\documentclass[12pt]{article}
\usepackage{amsmath,latexsym,epsfig,amssymb,eufrak}

\input eqalign.sty     

\newcommand{\dds}{\stackrel{\leftrightarrow}{D}}

\newcommand\be{\begin{equation}}
\newcommand\ee{\end{equation}}
\newcommand\bea{\begin{eqnarray}}
\newcommand\eea{\end{eqnarray}}

\def\ma[#1,#2,#3,#4]  {{\left( \matrix{ #1  & #2 \cr
                                        #3  & #4 \cr } \right)}}

\begin{document}

\title{
{\vspace{-1cm} \normalsize
\hfill \parbox{40mm}{CERN-TH/99-54}}\\
Renormalization group invariant average momentum of non-singlet parton
densities}
\author{
Marco Guagnelli$^1$,  
Karl Jansen$^{2,}$\footnote{Heisenberg Foundation Fellow}  
$\;$ and Roberto Petronzio$^{1}$  
\\
{\footnotesize $^1$ Dipartimento di Fisica, Universit\`a di Roma 
{\em Tor Vergata} }\\
{\footnotesize and INFN, Sezione di Roma II} \\
{\footnotesize Via della Ricerca Scientifica 1, 00133 Rome, Italy } \\
{\footnotesize 
$^2$ CERN, 1211 Geneva 23, Switzerland  
}
}
\maketitle

\begin{abstract}
We compute, within the Schr\"odinger functional scheme, a renormalization
group invariant renormalization constant for the first moment of the
non-singlet parton distribution function. The matching of the results of
our non-perturbative calculation with the ones from hadronic matrix
elements allows us to obtain eventually a renormalization group invariant
average momentum of non-singlet parton densities, which can be translated
into a preferred scheme at a specific scale.
\end{abstract}

\pagebreak

Physical quantities that need renormalization, such as the coupling constant,
the quark mass or the matrix elements of operators appearing in the Wilson
operator product expansion with a non-zero anomalous dimension, are
``running'' with the renormalization scale.  The choice of the scale is in
general motivated by the kinematics of the Green functions involving such
renormalized quantities, but the final physical predictions of the theory
without perturbative approximations are independent of such a
choice. This leads to the well-known renormalization group equations that
put the independence on a formal basis.  The redundancy in a
parametrisation of the theory in terms of renormalized quantities and the
relative renormalization scale can be avoided by considering
renormalization group invariant quantities, such as the $\Lambda$ parameter of
QCD or the renormalization group invariant quark masses (RGIM).  The advantage of
the latter choice in non-perturbative lattice determinations of the quark
mass has recently been stressed by the authors of ref.~\cite{qmass} where
an essential part of the RGIM programme was carried out.

In particular, the definition of the RGIM, which corresponds, roughly speaking,
to a running mass at infinite renormalization scale, is free of
the renormalization scheme dependence that usually affects quantities
renormalized (in a given scheme) at a fixed scale.  It can hence be evolved
back to an arbitrary finite scale in a preferred scheme.

The purpose of this letter is to present a similar calculation for the
operator that corresponds to the average momentum of non-singlet parton
densities.  A lattice --- perturbative and non-perturbative --- study of
the evolution of such an operator has been discussed in
refs.~\cite{roberto_pert} and ~\cite{paper_1} to which we address the
reader for more details about the calculation that we here only shortly
summarize as follows.  We calculate the renormalization constant of the
twist-two non-singlet operator for the first moment of the quark
parton distribution defined by:
\begin{equation}
{\cal{O}}_{\mu \nu}^{qNS} = 
\bigl( \frac{i}{2}\bigl)^{n-1} {\bar{\psi}}(x)
\gamma_{\{\mu} 
\dds_{\nu\}} 
\frac{\lambda^f}{2} \psi(x)\ -\ \mbox{trace terms}\; ,
\label{eq:twist_two_continuum}
\end{equation}
where $\left\{\dots\right\}$ means symmetrization of the indices. 
We remark that the technique discussed here can be extended to higher
moments in an anloguous way.
The basic ingredient for the reconstruction of the non-perturbative scale
dependence of the renormalization constants of the above operator is the
finite-size step scaling function $\sigma_Z$ defined by:
\begin{equation}\label{Zs}
Z(sL) = \sigma_Z(s, 
\bar{g}^2(L))Z(L)\; ,
\end{equation}
where $L$ is the physical length that plays the role of the renormalization
scale, $s$ parametrizes the step size of the change in the scale, and $Z$
is the renormalization constant of the operator, which is defined by:
\begin{equation}
{\cal O}^{R}(\mu) = Z(1/\mu)^{-1}{\cal O}^{\rm bare}(1/L)\; .
\end{equation}
$Z$ is
obtained from the Schr\"odinger Functional (SF) matrix element, $\langle\dots\rangle_{\rm SF}$, 
of the operator on a finite volume
$L^3T$, normalized to its tree level\footnote{In the following we choose 
$T = L$.}:
\begin{equation}
\langle{\cal O}^{\rm bare}(1/L)\rangle_{\rm SF} = Z(L) \langle{\cal O}^{\rm tree}\rangle_{\rm SF}\; .
\end{equation}
The renormalized operator then satisfies 
$\langle {\cal O}^R(\mu=1/L)\rangle_{\rm SF}=\langle {\cal O}^{\rm tree}\rangle_{\rm SF}$.
The framework of the Schr\"odinger Functional \cite{schrfunc,sint}, which describes the quantum
time evolution between two fixed classical gauge and fermion configurations,
defined at times $t=0$ and $t=T$, has been used extensively in the recent
literature \cite{letter,paper4,qmass} to calculate non-perturbative
renormalization constants of local operators.  Among the advantages of the
method, we only quote the possibility of performing the computations at
zero physical quark mass and of using non-local gauge-invariant sources for
the fermions without need of a gauge-fixing procedure. In our particular
case, we exploit both features. Our observable is defined
by~\cite{roberto_pert}:
\begin{equation}
Z =
\frac{f_2(x_0=L/4)}{\sqrt{f_1}}\left/\left(\frac{f_2(x_0=L/4)}{\sqrt{f_1}}\right)_{\rm
tree}\right.\; ,
\label{eq:obs}
\end{equation}
with $f_2$ given by
\begin{equation}
f_2(x_0) = -a^6\sum_{\bf{y},\bf{z}} \rm{e}^
{i\bf{p}(\bf{y}-\bf{z})}
\langle \frac{1}{4} \bar\psi(x) \gamma_{\{1} 
\dds_{2\}}\frac{1}{2} \tau^3 \psi(x) 
\bar\zeta({\bf{y}}) \Gamma \frac{1}{2} \tau^3 \zeta({\bf{z}})\rangle
\label{eq:f2}
\end{equation}
and $f_1$ by
\begin{equation}
f_1 = -a^{12}\sum_{\bf{y},\bf{z},\bf{v},\bf{w}}\langle
\bar\zeta'(\bf{v})\frac{\tau^3}{2}\zeta'(\bf{w})
\bar\zeta(\bf{x})\frac{\tau^3}{2}\zeta(\bf{y})\rangle\; ,
\label{eq:f1}
\end{equation}
where $\zeta=\delta /\delta\bar{\psi}_c$ and
$\bar\zeta=-\delta/\delta\psi_c$ are the derivatives with respect to the
two-component classical fermion fields ($\bar{\psi}_c$ and $\psi_c$,
respectively) at the boundary $x_0=0$, while $\zeta'$ and $\bar\zeta'$ are
the corresponding derivatives at the boundary $x_0=T$. The projection on
the classical components is achieved by the projector $P_{\pm}$ defined by
$\frac{1}{2}(1\pm\gamma_0)$. On the boundaries, the theory possesses only a
{\em global} gauge invariance that is preserved by the quantities defined
above.  The values of $x_0$ (set to $T/4$) and of the non-zero component of
the momentum $p_x$ (set to $2\pi$/$L$) are both scaled in units of $L$,
which therefore remains the only scale besides the lattice spacing $a$. The
quantity $f_1$ serves as a normalization factor that removes the wave
function renormalization constant of the $\zeta$ fields in order to isolate
the running associated with the operator in
eq.~(\ref{eq:twist_two_continuum}) only.

The determination of the step scaling function in the continuum has been
shown to be universal with respect to the lattice action used in
ref.~\cite{paper_2}.  From a fit to its dependence upon the running coupling
constant $\bar{g}^2$, renormalized in the SF scheme, we can extract the
following ``running'' step scaling function:

\begin{equation}
\sigma ( \mu/\mu_0,\bar{g}^2(\mu_0) ) = Z(1/\mu)/ Z(1/\mu_0)
\label{eq:step_running}
\end{equation}
i.e. the renormalization constant normalized to the one at a reference
scale $\mu_0$. 

The running  operator matrix element at the scale $\mu$, which we denote
generically by the symbol $O$, can be defined in terms
of the one at scale $\mu_0$ simply by:
\begin{equation}
  O^{ren}(\mu) = O^{ren}(\mu_0)\sigma ( \mu/\mu_0,\bar{g}^2(\mu_0) )\; .
\label{eq:operator_running}
\end{equation}
The scale dependence of the renormalized operator just reflects the one
of its renormalization constant governed by the equation:
\begin{equation}
\frac{{\rm d} Z(1/\mu)}{{\rm d}\log(\mu)} = Z(1/\mu)\cdot\gamma_O(g^2(\mu))\; ,
\label{eq:ren_g_Z}
\end{equation}
from which follows:
\begin{equation}
\frac{{\rm d}O^{ren}(\mu)}{{\rm d}\log(\mu)} = O^{ren}(\mu)\cdot\gamma_O(g^2(\mu))\; .
\label{eq:ren_g_O}
\end{equation}
Following ref.~\cite{qmass} but using a slightly different normalization
in taking out the factor of $2b_0$,
we define, for operators entering the
Wilson operator product expansion, a renormalization group invariant matrix
element:
\begin{equation}
  O^{ren}_{\rm INV} = O^{ren}(\mu)\cdot
                                   (\bar{g}^2)^{-\gamma_0/2b_0} \exp\left\{
    -\int_0^{\bar{g}} dg\left[\frac{\gamma(g)}{\beta(g)} - \frac{\gamma_0}{b_0g}\right]\right\}\; ,
\label{eq:O_inv}
\end{equation}
where for the anomalous dimension function $\gamma(g)$ and the
$\beta$-function the expressions up to three loops may be inserted
for values of $g$ small enough that perturbation theory can be trusted: 
\begin{equation} \label{gamma_3loop}
\gamma(g^2(\mu)) = \gamma_0 g^2(\mu) + \gamma_1 g^4(\mu) + \gamma_2 g^6(\mu),
\end{equation}
\begin{equation} \label{beta_3loop}
\beta(g^2(\mu)) = \beta_0 g^4(\mu) + \beta_1 g^6(\mu) + \beta_2 g^8(\mu).
\end{equation}
We note that for $\gamma(g)$ 
we know the effective three-loop term from our
non-perturbative computation of $\gamma_2$ \cite{paper_1},
while $\gamma_0$ and $\gamma_1$ are given
from perturbation theory. 
 
From eq.~(\ref{eq:step_running}) once the $O^{ren}(\mu_0)$ is known in some
scheme, for example the SF scheme we have described, we can obtain the
renormalization scheme invariant matrix element
by introducing an
``ultraviolet invariant'' running step scaling function 
\footnote{We remark that the invariance holds with
respect to a change of the ``ultraviolet'' scale $\mu$ and not 
of the ``infrared'' scale $\mu_0$.} 
defined by:
\begin{equation}
  \EuFrak{S}_{\rm INV}^{\rm UV}(\mu_0) =\sigma ( \mu/\mu_0,\bar{g}^2(\mu_0) )\cdot
                     (\bar{g}^2(\mu))^{-\gamma_0/2b_0} \exp\left\{
    -\int_0^{\bar{g}(\mu)} dg\left[\frac{\gamma(g)}{\beta(g)} - \frac{\gamma_0}{b_0g}\right]\right\}
\label{eq:step_running_inv}
\end{equation}
as follows:
\begin{equation}
  O^{ren}_{\rm INV} = O^{ren}(\mu_0)\cdot\EuFrak{S}_{\rm INV}^{\rm UV}(\mu_0)\; .
\label{eq:O_inv_2}
\end{equation}
The scale $\mu_0$ is in general a low-energy scale, where the hadronic
matrix element can be calculated without severe finite volume effects. In
our case, it can be identified with a low-energy scale at which 
the evolution of the renormalization constant can be started. In particular we shall fix
this scale to be $2L_{\rm max}$ as in ref.~\cite{qmass}. Recently, $L_{\rm max}$ 
has been computed in terms of the low energy reference quantity $r_0$ \cite{rainer} in
\cite{rainerlmax}. In order to ``step down'' from this scale, we will need the
step scaling function with $s=2$, i.e. starting from 
 $\bar{g}^2(L_{\rm max})=3.48$, our largest value of $\bar{g}^2$, we evolve 
with a step size of 2 until 
contact with perturbation theory can be made. 

In this paper, we calculate, as a first step towards the computation of the renormalization
group invariant matrix element, the quantity 
$\EuFrak{S}_{\rm INV}^{\rm UV}(\mu_0 = \frac{1}{2L_{\rm max}})$.
 Note that $\EuFrak{S}_{\rm INV}^{\rm UV}(\mu_0)$ still
depends on the reference scale $\mu_0$. The dependence on $\mu_0$ will only
disappear later, when it will be matched with the
proper hadronic matrix element, making $O_{\rm INV}^{ren}$ renomalization scheme
independent. 

In order to rely on the perturbative
expansion for the $\beta$- and $\gamma$-functions appearing in 
eq.~(\ref{eq:step_running_inv}), we had to extend the calculation of our
non-perturbative running to higher scales. We added four more 
values of $\bar{g}$ for
the step scaling function that now covers, in total, values of  
$\bar{g}^2(L)$ ranging from
$\bar{g}^2(L) = 3.48$ to $\bar{g}^2(L) = 0.8873$. 
For the results at the four lowest values of $\bar{g}$ 
we used the non-perturbatively improved clover action \cite{paper3}; 
in figs. 1 and 2 we
report the continuum extrapolation, for the values of $\bar{g}^2$ not
presented already in ref.~\cite{paper_2} of the
step scaling function of the quantities $f_1$ and $f_2$ of eq.~(\ref{eq:obs})
($\sigma_{\bar{Z}}$ and $\sigma_{f_1}$, respectively, see
\cite{paper_1}): at smaller length scales
the effects of lattice artefacts for $\sigma_{\bar{Z}}$ are 
progressively reduced and the extrapolations become flatter.

\begin{figure}
\vspace{0.0cm}
\begin{center}
\psfig{file=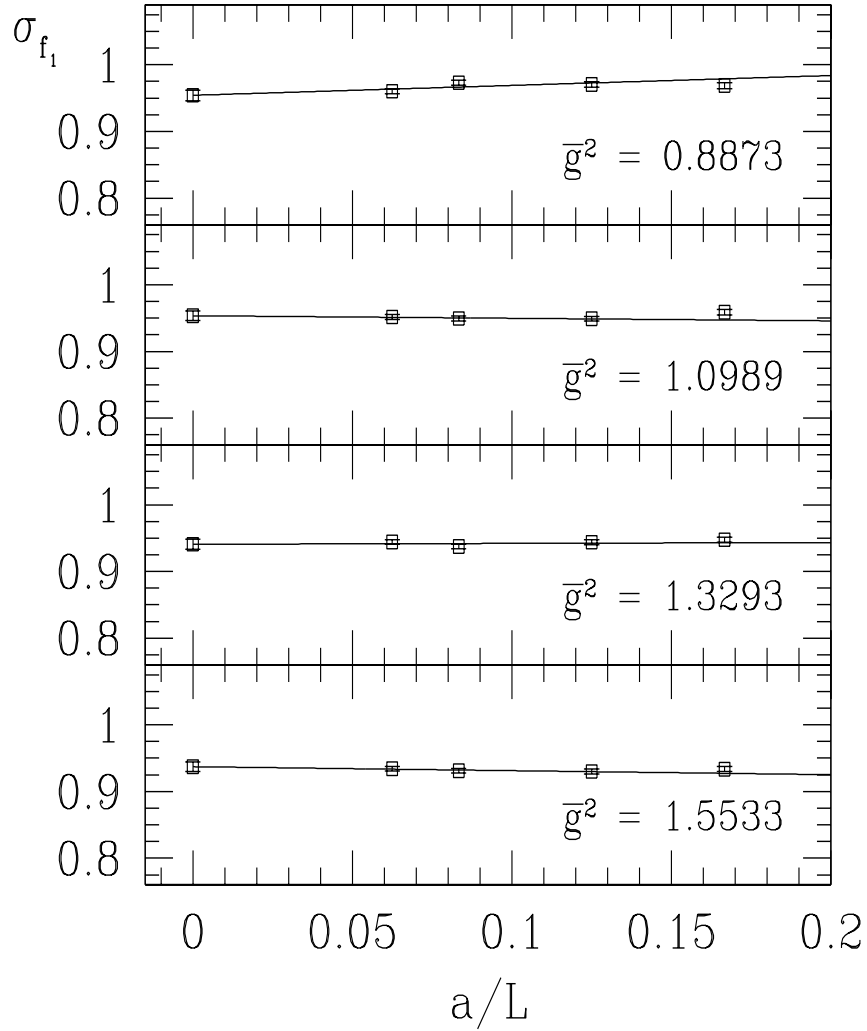, %
width=13cm,height=16cm}
\end{center}
\caption{ \label{fig:f1} Continuum extrapolation of
$\sigma_{f_1}$ using a linear fit to the three data points with smallest values of
$a/L$ for the
most perturbative values
of $\bar{g}^2$ we have used in our work, which are indicated in the figure.
}
\end{figure}

\begin{figure}
\vspace{0.0cm}
\begin{center}
\psfig{file=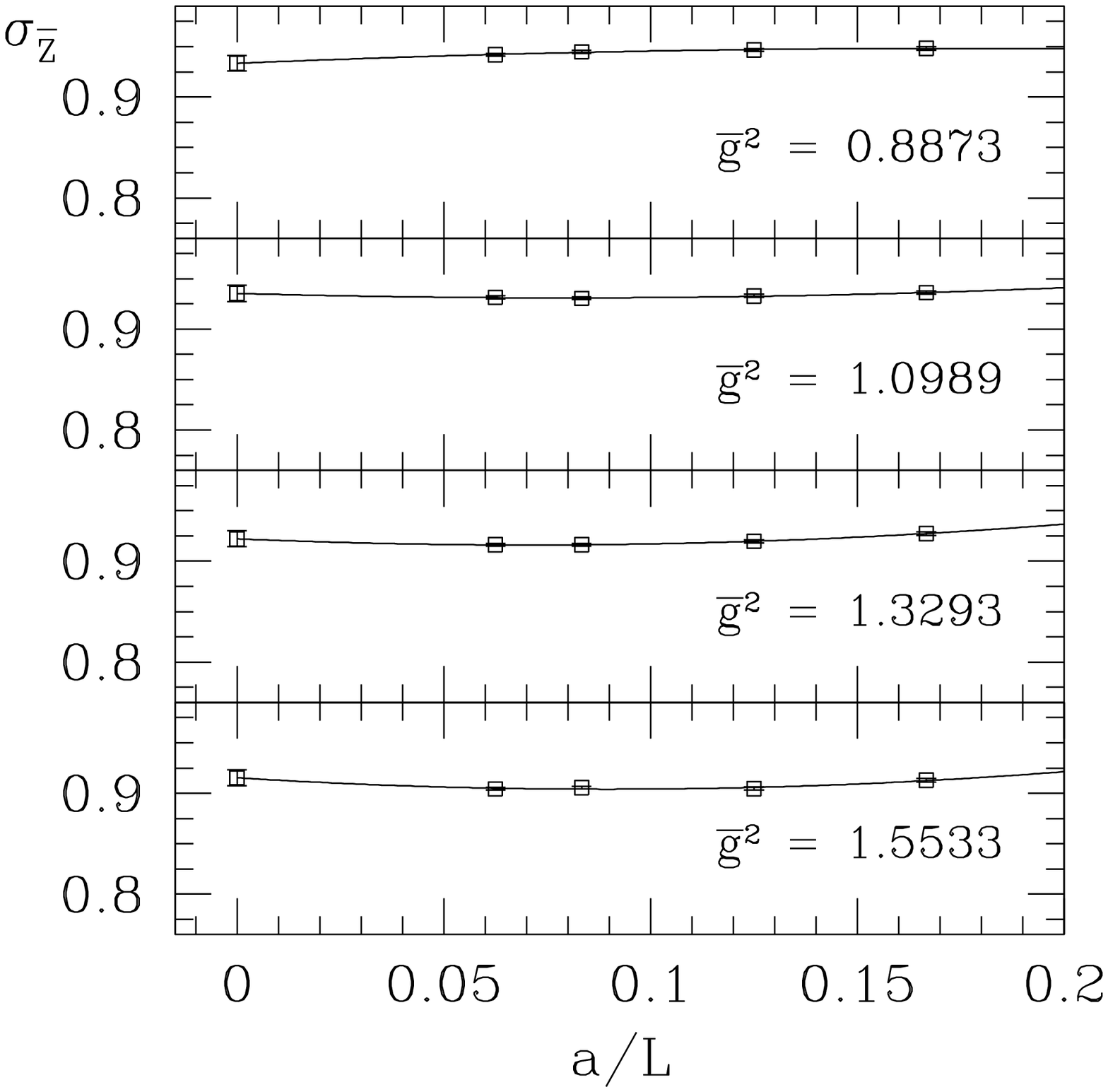, %
width=13cm,height=16cm}
\end{center}
\caption{ \label{fig:zbar_pert} Continuum extrapolation of
$\sigma_{\bar{Z}}$ using a quadratic fit to all four data points for the
most perturbative values of $\bar{g}^2$ we have used in our work, which
are indicated in the figure.}
\end{figure}

From the results for $\sigma_{Z}$ at the, in total, nine values of
$\bar{g}$, we can make a fit to the step scaling function as a function of
$\bar{g}^2(L)$. The results for $\sigma_{Z}$ at the five largest values of
$\bar{g}^2(L)$ are taken from the combined data presented in
\cite{paper_2}.  In ref.~\cite{paper_1} we have shown that, in the scheme we
adopted, the coefficient of the two-loop anomalous dimension is very
large, when compared for example to the one in the $\overline{\rm MS}$
scheme. We have also shown that this coefficient reduces by changing the 
expansion parameter,
i.e. by using $\bar{g}^2(L/4)$ instead of $\bar{g}^2(L)$.  The step scaling
function as a function of $\bar{g}^2(L/4)$ is well fitted numerically by
a polynomial in $\bar{g}^2(L/4)$ of the form:
\begin{equation}
\sigma(\bar{g}^2(L/4)) = 1 - \gamma_0\log(2)\bar{g}^2 + c_4\cdot \bar{g}^4 + c_6\cdot \bar{g}^6 + 
                                                                             c_8\cdot \bar{g}^8\; ,
\label{eq:step_fit}
\end{equation}
where $\gamma_0 = 4/(9\pi^2)$.
The final results stay
unchanged when we switch to a two-parameter fit that also gives a very good $\chi^2$.
We show our data for $\sigma_{Z}$ as a function of $\bar{g}^2(L/4)$ 
together with the fit of eq.~(\ref{eq:step_fit}) in fig.~3. 

\begin{table}[htbp]
  \begin{center}
    \leavevmode
    \begin{tabular}[]{|c|c|c|c|c|}
\hline
 $\mu/\mu_0$ & $\EuFrak{S}_{\rm INV}^{\rm UV}(\mu_0)$ &  $\EuFrak{S}_{\rm INV}^{\rm UV}(\mu_0)$ 
             & $\EuFrak{S}_{\rm INV}^{\rm UV}(\mu_0)$ &  $\EuFrak{S}_{\rm INV}^{\rm UV}(\mu_0)$ \\
 $\;$        &    $\bar{g}(L/4)$       &  $\bar{g}(L)$            
             &$\bar{g}(L/4)$           &  $\bar{g}(L)$            \\
 $\;$        & $\gamma_O$ 2-loop     & $\gamma_O$ 2-loop      
             & $\gamma_O$ 3-loop     & $\gamma_O$ 3-loop      \\
\hline\hline
$2^1$ &  $1.09(1)$ & $ 1.33(1)$ &  $1.16(1)$  &  $1.18(1)$  \\               
$2^2$ &  $1.10(2)$ & $ 1.24(2)$ &  $1.15(2)$  &  $1.16(2)$  \\  
$2^3$ &  $1.11(2)$ & $ 1.20(2)$ &  $1.14(2)$  &  $1.15(2)$  \\
$2^4$ &  $1.11(3)$ & $ 1.18(2)$ &  $1.14(3)$  &  $1.14(2)$  \\
$2^5$ &  $1.11(3)$ & $ 1.16(3)$ &  $1.13(3)$  &  $1.14(3)$  \\
$2^6$ &  $1.11(4)$ & $ 1.15(3)$ &  $1.13(4)$  &  $1.13(3)$  \\
$2^7$ &  $1.11(4)$ & $ 1.14(3)$ &  $1.12(4)$  &  $1.13(3)$  \\
$2^8$ &  $1.10(5)$ & $ 1.14(3)$ &  $1.12(5)$  &  $1.13(3)$  \\
$2^9$ &  $1.10(5)$ & $ 1.13(3)$ &  $1.11(5)$  &  $1.12(3)$  \\    
\hline
    \end{tabular}
    \caption{The values for $\EuFrak{S}_{\rm INV}^{\rm UV}(\mu_0)$ when different scales 
             $\mu$ are taken for matching with perturbation theory.}
    \label{tab:sigmamu0}
  \end{center}
\end{table}

From this fit we can construct the running step scaling function of
eq.~(\ref{eq:step_running_inv}) with $\mu_0=(2L_{\rm max})^{-1}$. The result is shown in fig.~4, where
we have used the two-loop expression for $\gamma(g)$ and the
3-loop expression for $\beta(g)$.  By using
eq.~(\ref{eq:step_running_inv}) we can finally estimate the value of
$\EuFrak{S}_{\rm INV}^{\rm UV}(\mu_0)$: in the second column of table~1 we report the values
of $\EuFrak{S}_{\rm INV}^{\rm UV}(\mu_0)$ as a function of the scale $\mu$: for large values
of $\mu$ the function, within the errors, approaches a plateau.  We make a
fit to a constant for the results ranging from $\mu/\mu_0 = 2^5$ to
$\mu/\mu_0 = 2^9$, and we finally quote:
\begin{equation}
 \EuFrak{S}_{\rm INV}^{\rm UV}(\mu_0=(2L_{\rm max})^{-1}) = 1.11(2)\; .
\end{equation}

\begin{figure}
\vspace{0.0cm}
\begin{center}
\psfig{file=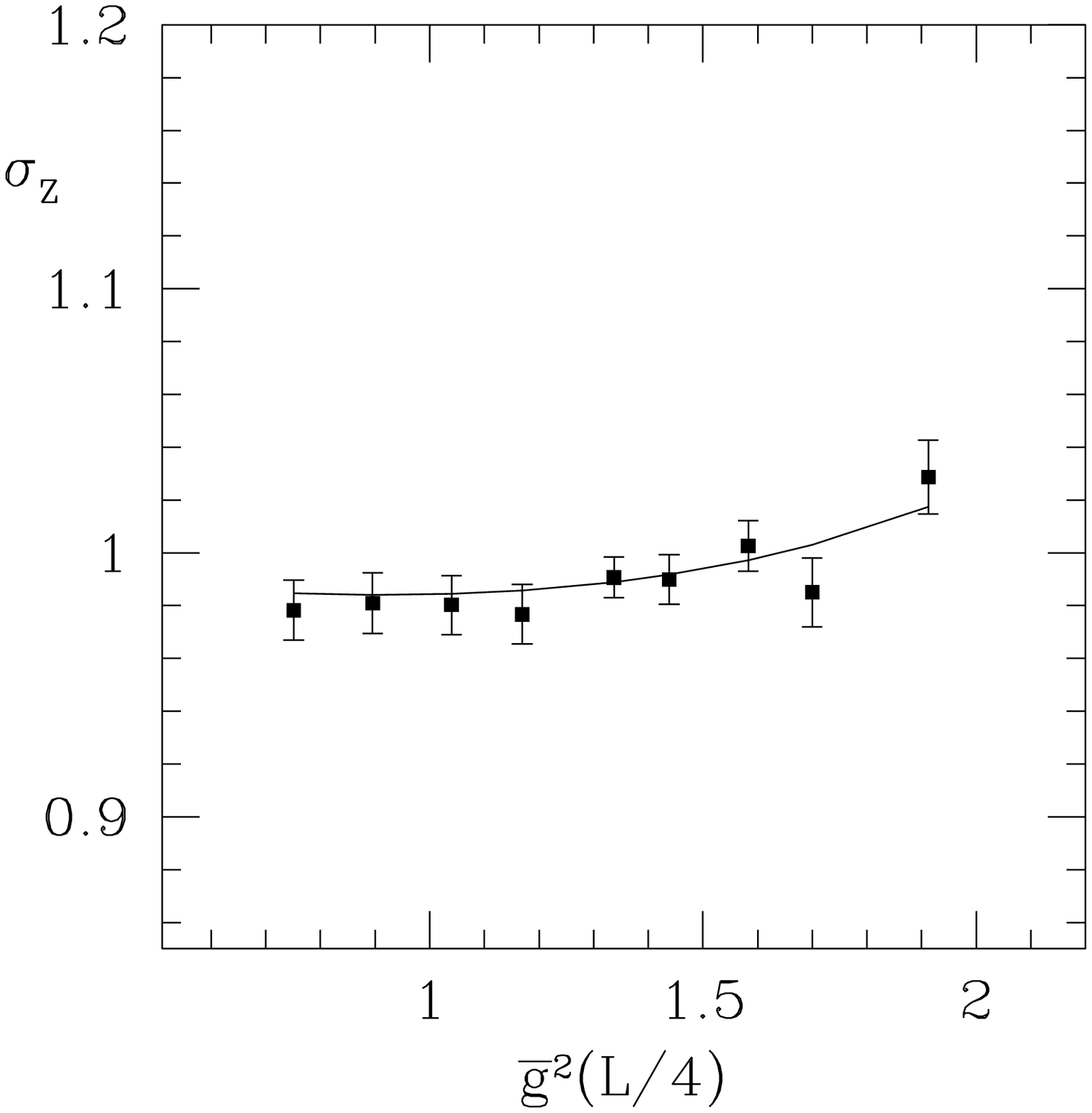, %
width=12cm,height=12cm}
\end{center}
\caption{ \label{fig:sgma_z} Our fit to the step scaling function. 
}
\end{figure}
%
\begin{figure}
\vspace{0.0cm}
\begin{center}
\psfig{file=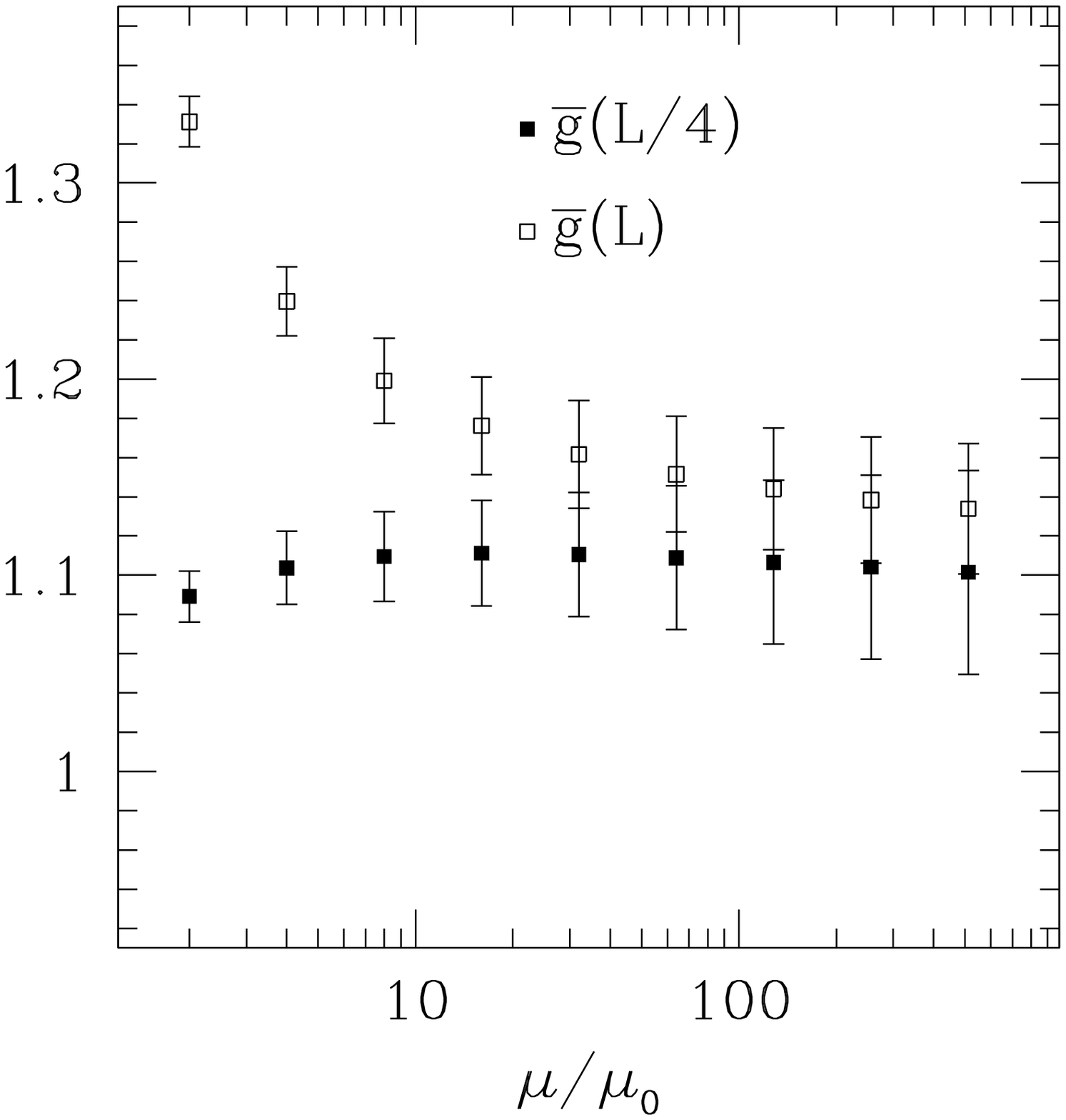, %
width=12cm,height=12cm}
\end{center}
\caption{ \label{fig:rgi} $\EuFrak{S}_{\rm INV}^{\rm UV}(\mu_0)$ as a function of the
scale $\mu$ normalized to our reference scale $\mu_0=(2L_{\rm max})^{-1}$. 
}
\end{figure}
The invariant step scaling function is still scheme-dependent, because of
the presence of the reference scale $\mu_0$. This will be cancelled only in
the combination that defines the invariant matrix element.  However, at
fixed $\mu_0$, it should be independent of the choice of $\bar{g}^2(L/4)$
or of $\bar{g}^2(L)$ in the fit to the
step scaling function. We therefore repeated the whole procedure described
above by fitting the step scaling function as a function of $\bar{g}^2(L)$ and by
using the correspondingly modified gamma function to two loops.  The
results are given in the third column of table~1.  They are fully
compatible with those obtained from the case $L/4$, although the plateau
starts at higher energies, as expected.  We report the comparison of both
cases also in fig.~\ref{fig:rgi}. In the fourth and fifth column of table
1 we report the result for the case ``$L/4$'' and ``$L$'' respectively, after
including our estimate of the three-loop anomalous dimensions for the two
cases, determined in~\cite{paper_1,paper_2}.  Not surprisingly, the two
cases get close to each other more precociously.  An estimate of the
renormalization group invariant yields
$\EuFrak{S}_{\rm INV}^{\rm UV}(\mu_0=\frac{1}{2L_{\rm max}})=1.14(2)$, again consistent with our
earlier results using $\bar{g}(L/4)$ as expansion parameter. 

Matching the results of this paper with a non-perturbative
calculation of the hadronic matrix element, in the continuum, in the same
scheme and at the same reference energy scale, leads to the definition of a
renormalization group invariant matrix element that can be confronted with
experiment at any scale and in a preferred scheme. Such a calculation is in
progress.

\section*{Acknowledgement}
We thank Martin L\"uscher for discussions and 
helpful comments. 


\end{document}